# Graphene oxide for enhanced nonlinear optics in low and high index contrast waveguides and nanowires

Yunyi Yang,[1,a] Jiayang Wu,[1,a] Xingyuan Xu,[1] Yao Liang,[1] Sai T. Chu,[2] Brent E. Little,[3] Roberto Morandotti,[4,5,6] Baohua Jia,[1,b] and David J. Moss[1,b]

[1]*Centre for Micro-Photonics, Swinburne University of Technology, Hawthorn, VIC 3122, Australia*

[2]*Department of Physics and Material Science, City University of Hong Kong, Tat Chee Avenue, Hong Kong, China.*

[3]*Xi'an Institute of Optics and Precision Mechanics Precision Mechanics of CAS, Xi'an, 710119, China.*

[4]*INSR-Énergie, Matériaux et Télécommunications, 1650 Boulevard Lionel-Boulet, Varennes, Québec, J3X 1S2, Canada.*

[5]*National Research University of Information Technologies, Mechanics and Optics, St. Petersburg, 197101, Russia.*

[6]*Institute of Fundamental and Frontier Sciences, University of Electronic Science and Technology of China, Chengdu 610054, China.*

We demonstrate enhanced four-wave mixing (FWM) in doped silica waveguides integrated with graphene oxide (GO) layers. Owing to strong mode overlap between the integrated waveguides and GO films that have a high Kerr nonlinearity and low loss, the FWM efficiency of the hybrid integrated waveguides is significantly improved. We perform FWM measurements for different pump powers, wavelength detuning, GO coating lengths, and number of GO layers. Our experimental results show good agreement with theory, achieving up to ~9.5-dB enhancement in the FWM conversion efficiency for a 1.5-cm-long waveguide integrated with 2 layers of GO. We show theoretically that for different waveguide geometries an enhancement in FWM efficiency of ~ 20 dB can be obtained in the doped silica waveguides, and more than 30 dB in silicon nanowires and slot waveguides. This demonstrates the effectiveness of introducing GO films into integrated photonic devices in order to enhance the performance of nonlinear optical processes.

## I. INTRODUCTION

All-optical integrated photonic devices offer competitive solutions to achieve on-chip signal processing without the need for complex and inefficient optical-electrical-optical (O-E-O) conversion.[1] By directly processing signals in the optical domain, these devices can reduce power consumption and increase the bandwidth of optical telecommunications systems, with the added benefits of a compact footprint, high stability, mass-producibility, and the potential to significantly reduce cost.[2,3] Four wave mixing (FWM), as an important nonlinear optical process, has been widely used to for all-optical signal processing functions such as wavelength conversion,[4,5] optical logic gates,[6,7] optical comb generation,[8,9] quantum entanglement,[10,11] and more.[12,13] Efficient FWM in the telecommunications band has been demonstrated using silica fibers,[14,15] III/V semiconductor optical amplifiers (SOAs),[16,17] integrated photonic devices based on silicon[12,18] and other complementary metal–oxide–semiconductor (CMOS) compatible platforms,[19,20] polymer composites,[21,22] chalcogenide devices,[23,24] and others.[25,26]

Although silicon has been a leading platform for integrated photonic devices, its high two-photon absorption (TPA) at near-infrared wavelengths is a fundamental limitation for the performance of nonlinear devices in the telecommunications band.[27] Hence, the quest for high-performance integrated platforms for nonlinear optics has motivated the development of other CMOS compatible platforms such as silicon nitride and high-index doped silica glass.[3] Benefiting from extraordinarily low linear and nonlinear loss, high-index doped silica glass has been a successful integrated platform for nonlinear photonic devices.[9–11, 28–32] Nevertheless, its relatively low Kerr nonlinearity as compared with silicon and silicon nitride[3] limits its performance for nonlinear optical processes. The use of highly nonlinear materials that can be integrated on chip could overcome these limitations. Owing to its ease of preparation as well as the tunability of its material properties, graphene oxide (GO) has become a highly promising member of the graphene family.[33] Previously,[34,35] we reported GO films with a giant Kerr nonlinear response of 4 to 5 orders of magnitude higher than that of high-index doped silica glass. Moreover, as compared with graphene, GO has much lower loss and larger bandgap (2.4~3.1 eV) [36,37] which yields low TPA in the telecommunications band. It also offers better capability for large-scale fabrication,[38] critical for the practical implementation of high-performance nonlinear photonic devices.

---

[a]These authors contribute equally to this paper.

[b]Author to whom correspondence should be addressed. Electronic mail: bjia@ swin.edu.au, dmoss@swin.edu.au.

In this paper, we report significantly improved FWM performance for high-index doped silica glass waveguides by integrating them with GO films. By designing and fabricating integrated waveguides with a strong mode overlap with the GO film, we achieve enhanced FWM efficiency in the GO hybrid integrated waveguides, with an experimental conversion efficiency (CE) enhancement of ~9.5 dB for a 1.5-cm-long integrated waveguide with 2 layers of GO. FWM measurements are also performed for different pump powers, wavelength detuning, GO coating lengths, and number of GO layers, with good agreement between experiment and theory. Finally, we show theoretically that by optimizing the device parameters for the low index contrast waveguides studied here, an enhancement in the FWM efficiency of ~20dB is possible, and for silicon nanowires and slot waveguides, more than 30 dB enhancement is achievable. These results confirm the improved FWM performance of integrated photonic devices incorporated with GO.

**II. DEVICE FABRICATION AND CHARACTERIZATION**

Figure 1(a) shows the GO-coated integrated waveguides made from high-index doped silica glass,[3] with a cross section of 2 μm × 1.5 μm. The integrated waveguide is surrounded by silica except that the upper cladding is removed to enable coating the waveguide with GO films. The GO films, with a thickness of about 2 nm per layer, were introduced on the top of the integrated waveguide in order to introduce light-material interaction with the evanescent field leaking from the integrated waveguide. The Kerr coefficient of GO is on the order of $10^{-15}$~$10^{-14}$ m$^2$/W,[34,35] which is slightly lower than that of graphene (~$10^{-13}$ m$^2$/W),[25,35,39] but still orders of magnitude higher than that of high-index doped silica glass (~$10^{-19}$ m$^2$/W) and silica (~$10^{-20}$ m$^2$/W).[3] The waveguides were fabricated via CMOS compatible processes.[40,41] First, high-index doped silica glass films ($n$ = ~1.60 at 1550 nm) were deposited using standard plasma enhanced chemical vapour deposition (PECVD), then patterned using deep UV photo-lithography and etched via reactive ion etching (RIE) to form waveguides with exceptionally low surface roughness. After that, silica glass ($n$ = ~1.44 at 1550 nm) was deposited via PECVD and the upper cladding of the integrated waveguides was removed by chemical mechanical polishing (CMP). Finally, the GO film was coated on the top surface of the chip by a solution-based method that yields layer-by-layer deposition of GO films, as reported previously.[42] As compared with graphene that is typically prepared by mechanical exfoliation or chemical vapour deposition, both needing sophisticated transfer processes, GO can be directly coated on dielectric substrates (e.g., silicon and silica wafers) via a solution-based approach.[42,43] This approach is capable of coating large areas (e.g., a 4-inch wafer) with relatively few defects, which is critical for the fabrication of large-scale integrated devices.

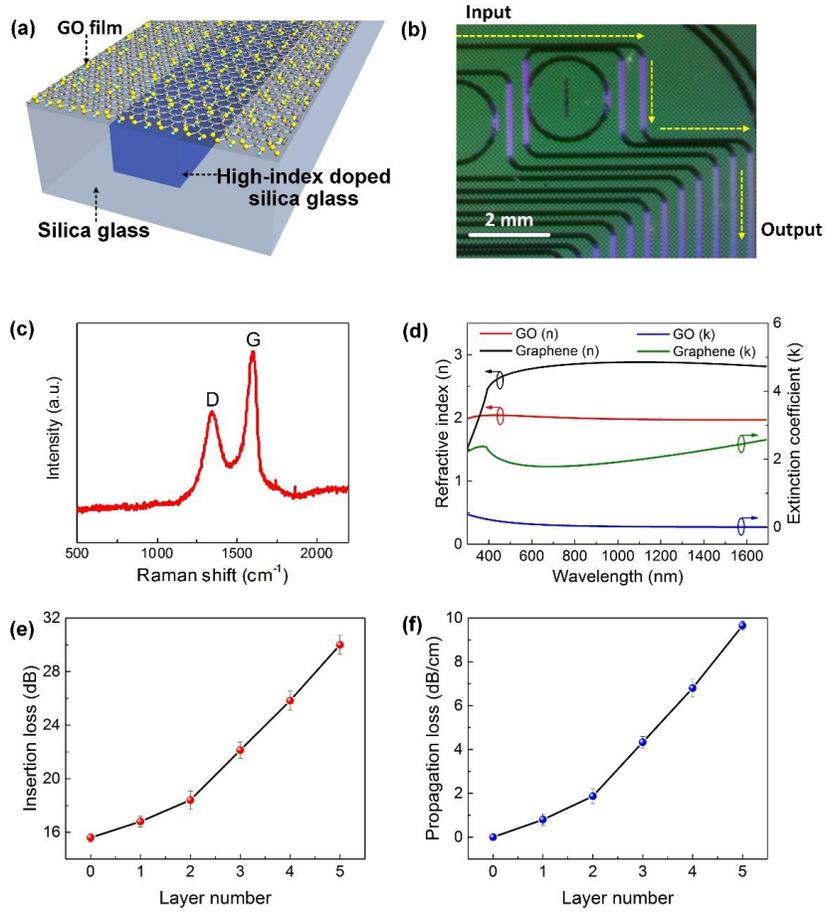

Fig. 1. (a) Schematic illustration of hybrid waveguides integrated with GO. (b) Image of the hybrid integrated waveguide with two layers of GO. (c) Raman spectra of GO on the integrated chip. (d) Measured refractive indices and extinction coefficients of GO and graphene. (e) Measured insertion loss of a 1.5-cm-long integrated waveguide with different numbers of GO layers. (f) Additional propagation loss of the integrated waveguide with different numbers of GO layers.

An image of the integrated waveguide incorporating two layers of GO is shown in Fig. 1(b), which illustrates that the morphology is good, leading to a high transmittance of the GO film on top of the integrated waveguide. The integration of GO onto the waveguide is confirmed by Raman spectroscopic measurements (Fig. 1(c)) that show the representative D (1345 cm$^{-1}$) and G (1590 cm$^{-1}$) peaks of GO.[34] Figure 1(d) shows the in-plane refractive index ($n$) and extinction coefficient ($k$) of the GO film including 5 layers of GO measured by spectral ellipsometry.[44] For comparison, the refractive index and extinction coefficient of graphene are also shown in Fig. 1(d). The GO film exhibits a high refractive index of ~ 2 in the telecommunications band. On the other hand, due to the existence of oxygen functional groups (OFGs), the GO films also exhibit an ultra-low extinction coefficient in the telecommunications band, which leads to much lower material absorption as compared with graphene. This property of GO could be of benefit for FWM devices, where low loss is always desired for improved efficiency. It should be noted that, unlike graphene where the bandgap is zero, GO has a distinct bandgap of 2.4 to 3.1 eV,[36,37] which results in low linear and nonlinear light absorption in spectral regions below the bandgap, in particular featuring greatly reduced TPA in the telecommunications band. This represents a significant advantage over graphene. However, GO still has significantly higher propagation loss than silicon or silica at 1.55 μm due to higher scattering loss and absorption by impurities and defects.

To couple light into the hybrid integrated waveguides we employed an 8-channel single-mode fibre (SMF) array for butt coupling. The propagation loss of the uncoated doped silica waveguide was ~24 dB/m, which was negligible for the length (~1.5 cm) used here, and so the total insertion loss for bare (uncoated) waveguides was dominated by the coupling loss, which was ~8 dB/ facet. This can be reduced to ~1.5 dB/facet by using mode convertors between SMF and the waveguides. We chose the transverse electric (TE) field polarization for the experiments because it supports in-plane interaction between the evanescent field and the thin GO film, which is much stronger than out-of-plane interaction due to the large optical anisotropy of 2D materials, as is the case for graphene [45,46] (see discussion in Section IV on polarization dependent issues). Figure 1(e) depicts the total insertion loss of the integrated waveguides with different numbers of GO layers measured with continuous-wave (CW) light at a wavelength of 1550 nm. The measured insertion loss did not show any significant variation with power (up to ~30 dBm) and wavelength (1500~1600 nm) of the CW light. We characterized four duplicate integrated waveguides with the same length of ~1.5 cm. The GO layers only affected the propagation (not coupling) loss, which is shown in Fig. 1(f) as a function of the number of layers. The overall propagation loss of the GO hybrid integrated waveguides was on the order of a few dB/cm, which is much lower than that of graphene hybrid integrated waveguides[47] and confirms the low material absorption of GO in the telecommunications band. Another interesting phenomenon is that the rate of increase in propagation loss with layer number increases (i.e., becomes super-linear) for higher numbers of GO layers. This might be attributed to interactions among the GO layers as well as possible imperfect contact between the multiple GO layers in practical devices.

## III. EXPERIMENT

We used the experimental setup shown in Fig. 2 to perform FWM measurements in the GO hybrid integrated waveguides. Two CW tunable lasers were separately amplified by erbium-doped fiber amplifiers (EDFAs) and used as the pump and signal sources, respectively. In each path, there was a polarization controller (PC) to ensure that the input light was TE-polarized. The pump and signals were combined by a 50:50 fiber coupler before being injected into the GO hybrid integrated waveguide. The signal output from the waveguide was sent to an optical spectrum analyser (OSA) with a variable optical attenuator (VOA) inserted before the OSA to prevent the high-power output from damaging it.

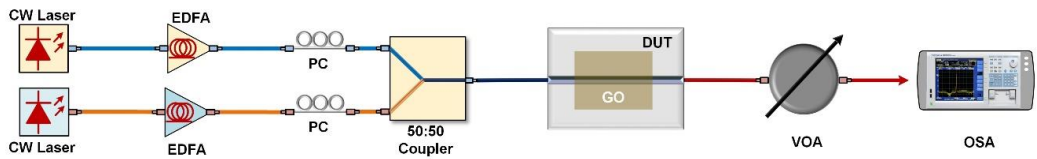

Fig. 2. Experimental setup for testing FWM in the GO hybrid integrated waveguide. EDFA: erbium-doped fiber amplifier. PC: polarization controller. DUT: device under test. VOA: variable optical attenuator. OSA: optical spectrum analyser.

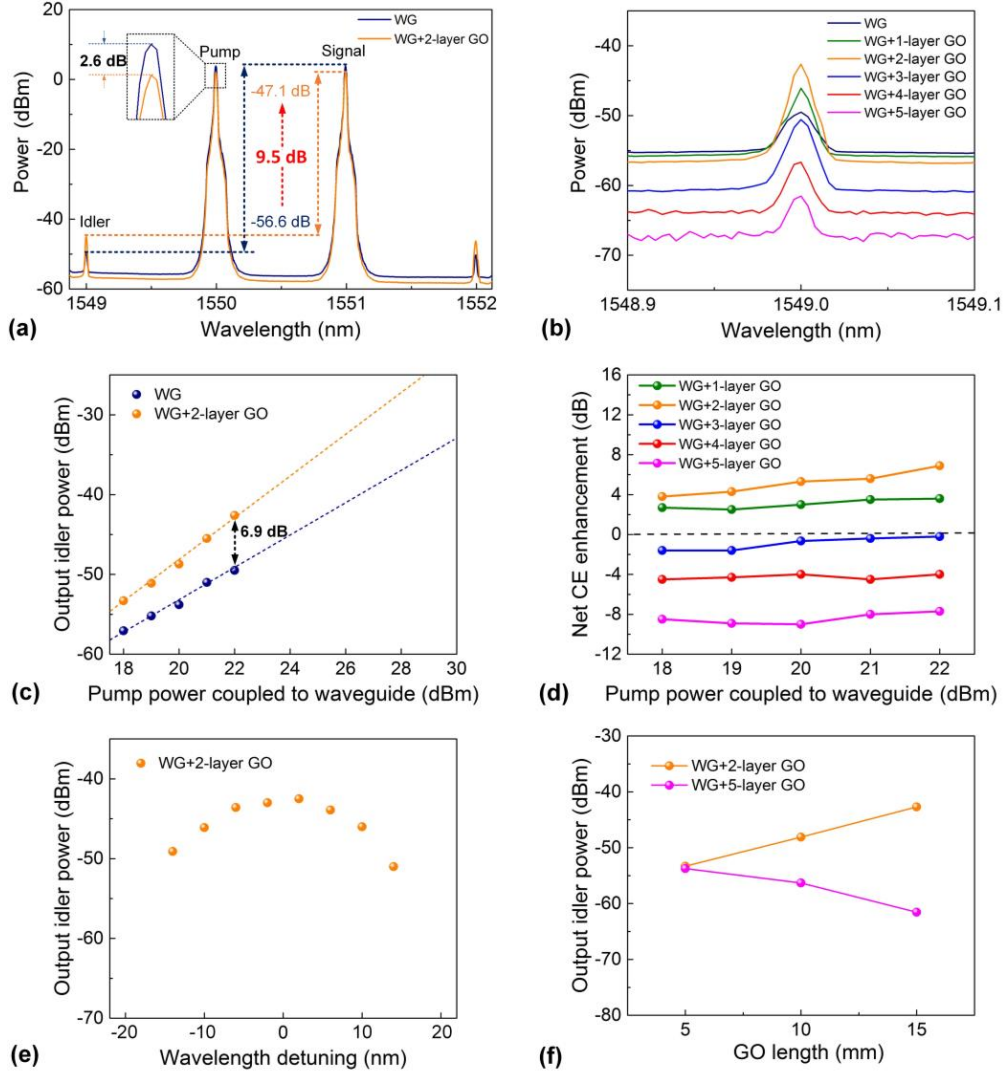

Fig. 3. FWM experimental results. (a) FWM spectra of the integrated waveguide without GO and with 2 layers of GO. (b) Zoom in spectra of the generated idlers after FWM in the waveguide with 0 to 5 layers of GO. (c) Output powers of idler for various pump powers coupled to the waveguide without GO and with 2 layers of GO. (d) Net CE enhancements for various pump powers coupled to the waveguide with 1 to 5 layers of GO. (e) Power variations of the output idler when the pump wavelength was fixed at 1550 nm and the signal wavelength was detuned around 1550 nm. (f) Output powers of idler for the waveguide with different coating lengths of GO. The GO coating length in (a), (b), (c), (d), (e) is ~1.5 cm. The pump power coupled to the waveguide in (a), (b), (e), (f) is ~22 dBm. WG: waveguide.

Figure 3 shows the FWM experimental results. The FWM spectra of a 1.5-cm-long integrated waveguide without GO and with 2 layers of GO are shown in Fig. 3(a). For comparison, we kept the same pump power of ~30 dBm before the input of the waveguide, which corresponded to ~22 dBm pump power coupled into the waveguide. It can be seen that although the hybrid integrated waveguide had additional propagation loss (~2.6 dB), it clearly shows enhanced idler output powers as compared with the same waveguide without GO. The CE (defined as the ratio of the output power of the idler to the output power of the signal, i.e., $P_{out,\ idler}/P_{out,}$ signal) of the integrated waveguide with and without GO were ~-47.1 dB and ~-56.6 dB, respectively, corresponding to a CE enhancement of ~9.5 dB for the hybrid integrated waveguide. After excluding the addtional propagation loss, the net CE enhancement (defined as the improvement of the output power of the idler for the same pump power coupled to the waveguide) is 6.9 dB. For the integrated waveguide coated with 1 to 5 layers of GO, zoom-in spectra of the generated idlers for the same pump power coupled to the waveguide (~22 dBm) are shown in Fig. 3(b). For the integrated waveguide coated with 1 and 2 layers of GO, there were positive net CE enhancements. When the number of GO layers was over 2, the net change in CE was negative. This is mainly due to the super-linear increase in propagation loss for increased numbers of GO layers as noted above. The output powers of the idler for various pump powers coupled to the waveguide without GO and with 2 layers of GO are shown in Fig. 3(c). The pump power was varied while the signal power remained constant. One can see that as the pump power increased, the idler power increased with no obvious saturation for both samples, which reflects the low nonlinear absorption of both the high-index doped silica glass as well as the GO layers in the telecommunications band. The slope of the curve for the uncoated waveguide is about 2, as expected from classical FWM theory, while the slope of the GO hybrid waveguide is slightly larger than 2. This indicates that the material properties of the GO film (e.g. $n_2$ and absorption) could be a function of pump power, as noted previously.[34,48] The physics of this phenomenon is the subject of on-going research. The net CE enhancement for various pump powers coupled to the waveguide with 1 to 5 layers of GO is shown in Fig. 3(d). There were positive net CE enhancements for all pump powers when the waveguide was coated with either 1 or 2 layers of GO. There was a slight increase in CE enhancement for high pump powers. This may also be caused by the slightly different material properties of the GO films at high light powers. The variation in idler power when the pump wavelength was fixed at 1550 nm and the signal wavelength detuned around 1550 nm is shown in Fig. 3(e). There is obvious power degradation of the output idler when the wavelength detuning is beyond ±10 nm. When the wavelength detuning was beyond ±15 nm, we could not observe any idler above the noise floor. The output powers of idler light for a 1.5-cm-long integrated waveguide coated with different lengths of GO are depicted in Fig. 3(f). We used three GO coating lengths of ~0.5 cm, ~1.0 cm, and ~1.5 cm. For the waveguide coated with 2 layers of GO, the output idler power increased with the GO length, whereas for the waveguide coated with 5 layers of GO, there was an opposite trend. This phenomenon further confirms the trade-off between FWM enhancement and propagation loss in these hybrid integrated waveguides.

## IV. THEORY

We used the theory in Refs. 18, 49, and 50 to model the FWM performance of the GO hybrid integrated waveguides. First, as is commonly done,[27,51,52] we assume that:

$$\chi^{(3)} (\omega_{idler}; \omega_{pump}, \omega_{pump}, -\omega_{signal}) \approx \chi^{(3)} (\omega; \omega, \omega, -\omega), \quad (1)$$

$$n_2 (\omega) = 3 \cdot \mathrm{Re}\,[\chi^{(3)}(\omega)] / (4cn_0^2\varepsilon_0), \quad (2)$$

where $\chi^{(3)}$ and $n_2$ are the third-order susceptibility and the Kerr nonlinear coefficient, respectively. $2\omega_{pump} = \omega_{idler} + \omega_{signal}$, with $\omega_{pump}$, $\omega_{signal}$, and $\omega_{idler}$ denote the angular frequencies of the pump, signal, and idler, respectively, $c$ is the speed of light in vacuum, $n_0$ is the linear refractive index, and $\varepsilon_0$ is the vacuum permittivity. Note that although Eq. (1) is often considered for FWM, it is only valid in the regime close to degeneracy where the three FWM frequencies (pump, signal, idler) are close together compared with any variation in $n_2$ arising from the dispersion in $n_2$. Since $n_2$ is expected[51] to vary significantly near the material bandgap, there is no guarantee that Eqs. (1) and (2) hold in cases where the material bandgap is comparable to, or even smaller than, the photon energies being employed. This particularly applies to graphene, for example. Further, Eq. (1) also ignores the 4$^{th}$ rank tensor nature of $\chi^{(3)}$. It is well known[52–54] that even for cubic materials such as silicon, the tensor elements for $\chi^{(3)}$ are not all equal, resulting in a variation in the nonlinear efficiency of up to 50% or more as a function of orientation in silicon or germanium. While the doped silica waveguides studied here are isotropic, it is not only possible but highly likely that the nonlinearity of the thin GO films is indeed highly anisotropic. This effect is in addition to any variations in the waveguide nonlinear parameter $\gamma$ (Eq. (6)) arising from differences in the mode overlap with the GO films between the two polarizations. In our experiments we restricted the polarization to TE and so effectively we measure the $n_2$ that applies to this polarization. We note that previous measurements of $n_2$ in these films were performed in transmission on broad area films and so corresponded to TE polarization in our experiments. This topic (the anisotropy of nonlinear processes in GO films) will be the subject of our future work.

Considering the linear loss as well as the nonlinear loss induced by TPA, and self / cross phase modulation, the coupled differential equations for the degenerate FWM processes can be expressed as:[49,50]

$$\frac{dA_p(z)}{dz} = \left[-\frac{\alpha_p}{2} - \frac{\alpha_{TPA}}{2}|A_p(z)|^2\right]A_p(z) + j\gamma_p\left[|A_p(z)|^2 + 2|A_s(z)|^2 + 2|A_i(z)|^2\right]A_p(z) + j2\gamma_p A_p^*(z)A_s(z)A_i(z)exp(j\Delta\beta z), \quad (3)$$

$$\frac{dA_s(z)}{dz} = \left[-\frac{\alpha_s}{2} - \alpha_{TPA}|A_p(z)|^2\right]A_s(z) + j\gamma_s\left[|A_s(z)|^2 + 2|A_p(z)|^2 + 2|A_i(z)|^2\right]A_s(z) + j\gamma_s A_i^*(z)A_p^2(z)exp(-j\Delta\beta z), \quad (4)$$

$$\frac{dA_i(z)}{dz} = \left[-\frac{\alpha_i}{2} - \alpha_{TPA}|A_p(z)|^2\right]A_i(z) + j\gamma_i\left[|A_i(z)|^2 + 2|A_p(z)|^2 + 2|A_s(z)|^2\right]A_i(z) + j\gamma_i A_s^*(z)A_p^2(z)exp(-j\Delta\beta z). \quad (5)$$

where $A_{p,s,i}$ are the amplitudes of the pump, signal, and idler waves along the z axis, which we assume as the light propagation direction, $\alpha_{p,s,i}$ are the power linear loss factors, $\alpha_{TPA}$ is the power loss factor induced by TPA, $\Delta\beta = \beta_s + \beta_i - 2\beta_p$ is the linear phase mismatch, with $\beta_{p,s,i}$ denoting the propagation constants of the pump, signal, and idler waves, and $\gamma_{p,s,i}$ are the nonlinear parameters of the hybrid waveguide.

According to Ref. [18], the general expression for $\gamma$ is:

$$\gamma = \frac{3\omega_p\varepsilon_0}{16P_\mu^2}\iint_D \left\{\tilde{\chi}^{(3)}\left(\omega_p;\omega_p,\omega_p,-\omega_p\right) \vdots \xi_\mu(\omega_p)\xi_\mu(\omega_p)\xi_\mu^*(\omega_p)\right\} \cdot \xi_\mu^*(\omega_p)dxdy, \quad (6)$$

where $\omega_p = 2\pi c/\lambda$ is the angular frequency of the optical pump, $\lambda$ the pump wavelength, $D$ is the integral domain over the material regions with the fields, $\xi_\mu$ is the electric field vectors of the waveguide mode $\mu$, and $P_\mu$ is the power normalization constant given by:

$$P_\mu = \frac{1}{2}\iint_D Re\left\{\xi_\mu(x,y)\times\mathcal{H}_\mu^*(x,y)\right\}\vec{e_z}\,dxdy = \frac{1}{2Z_0}\iint_D n_0(x,y)\xi_\mu^2(x,y)dxdy, \quad (7)$$

where $\mathcal{H}_\mu$ is the magnetic field vectors of the waveguide mode $\mu$, $\vec{e_z}$ is the unit vector pointing in positive z-direction, $Z_0 = \sqrt{\mu_0/\varepsilon_0}$ ($\mu_0$ is the vacuum permeability), and $n_0(x, y)$ the linear refractive index profile over the waveguide cross section. The factor $n_0(x, y) / Z_0$ is introduced to compensate for the difference between the amplitudes of the electric and magnetic fields. Although $\tilde{\chi}^{(3)}$ is a 4$^{th}$ rank tensor, in this work we approximate it with a scalar: $\tilde{\chi}^{(3)} \vdots \xi_\mu\xi_\mu\xi_\mu^* = \chi^{(3)}|\xi_\mu|^2\xi_\mu$. For materials with low TPA, $\chi^{(3)} \approx$ Re $[\chi^{(3)}]$, and so

using Eq. (2) one obtains:

$$\tilde{\chi}^{(3)} \vdots \xi_\mu \xi_\mu \xi_\mu^* \approx \frac{4n_0^2(x,y)n_2(x,y)}{3z_0}\left|\xi_\mu\right|^2 \xi_\mu. \tag{8}$$

Substituting Eqs. (7) and (8) in Eq. (6) yields:

$$\gamma = \frac{2\pi}{\lambda} \frac{\iint_D n_0^2(x,y)n_2(x,y)\xi_\mu^4(x,y)\mathrm{d}x\mathrm{d}y}{\left[\iint_D n_0(x,y)\xi_\mu^2(x,y)\mathrm{d}x\mathrm{d}y\right]^2}. \tag{9}$$

Since the time-averaged Poynting vector $S_z$ satisfies $S_Z \propto \left|\xi_\mu\right|^2$, Eq. (9) can be reduced to:

$$\gamma = \frac{2\pi}{\lambda} \frac{\iint_D n_0^2(x,y)n_2(x,y)S_z^2\mathrm{d}x\mathrm{d}y}{\left[\iint_D n_0(x,y)S_z\mathrm{d}x\mathrm{d}y\right]^2}. \tag{10}$$

In Eq. (10), we see that $\gamma$ is an effective nonlinear parameter weighted by both the linear refractive index $n_0$ $(x, y)$ and the nonlinear coefficient $n_2$ in the different material regions.

We note that a commonly used expression [39, 55] for $\gamma$ is

$$\gamma = \frac{2\pi}{\lambda} \frac{\iint_D n_2(x,y)S_z^2\mathrm{d}x\mathrm{d}y}{\left[\iint_D S_z\mathrm{d}x\mathrm{d}y\right]^2}. \tag{11}$$

which is similar to Eq. (10) except without the weighting factor $n_0(x, y)$. For low-index-contrast waveguides such as fibers, $n_0(x, y)$ is approximately constant and so Eq. (11) is a reasonable approximation to Eq. (10). For high-index-contrast hybrid waveguides, on the other hand, particularly for the silicon waveguides studied here, Eq. (10) is more accurate. We note that a different model has been used for graphene hybrid waveguides,[19] based on surface conductivity with $\sigma^{(3)} = -i\varepsilon_0\omega_p\chi^{(3)}$. Since our GO layers are much thicker (~2 nm / layer) than a monolayer of graphene (~0.35 nm), we believe that the surface conductivity $\sigma^{(3)}$ approach is not valid. We calculated $\gamma$ for all waveguides using Eq. (10) in conjunction with commercial software (COMSOL Multiphysics®). We considered the GO layers as a thin dielectric film, and meshed the region around the GO film with a super fine resolution of 0.2 nm to make the calculations more accurate.

By numerically solving Eqs. (3) – (5), we obtain the FWM efficiency:

$$\eta \text{ (dB)} = 10 \times \log_{10}[|A_i(L)|^2/|A_s(0)|^2], \tag{12}$$

where $L$ is the length of the waveguide. Note that $\eta$ is the ratio of the output power of idler light to the input power of signal light, i.e., $P_{out,\,idler}/P_{in,\,signal}$, which is slightly different from the definition of CE in section III. It should be noted that the effect of phase matching is negligible in our experiments since $L\beta_2\Delta\omega^2 \ll 1$,[19] where $L \approx 1.5$ cm is the waveguide length, $\beta_2 \approx -3.9 \times 10^{-26}$ s$^2$ m$^{-1}$ is the group velocity dispersion of the GO hybrid waveguide, and $\Delta\omega < 1.5 \times 10^{13}$ rad/s is the angular frequency detuning range.

## V. RESULTS AND DISCUSSION

The calculated TE mode profile of the hybrid doped silica integrated waveguide with 2 layers of GO is presented in Fig. 4(a). Figures 4(b) – (d) show the measured and fit FWM efficiency ($\eta$) for different pump powers, wavelength detuning, and GO lengths. In the calculations, we used the pump power coupled to the waveguide in our FWM experiment as the input pump power and assumed that the TPA in GO, silica, and doped silica is negligible. We also used the experimentally measured values in Fig. 1(d) and Fig. 2(f) for the dispersion and loss of GO. By using the theory in Section IV to fit our experimental results, we obtained a $\gamma$ for the bare waveguide and the waveguide with 2 layers of GO of 0.28 $W^{-1}m^{-1}$ and 0.90 $W^{-1}m^{-1}$, respectively. We then obtained $n_2$ of the GO film from $\gamma$ using Eq. (10) and the results are shown in Table I along with other results from the literature.

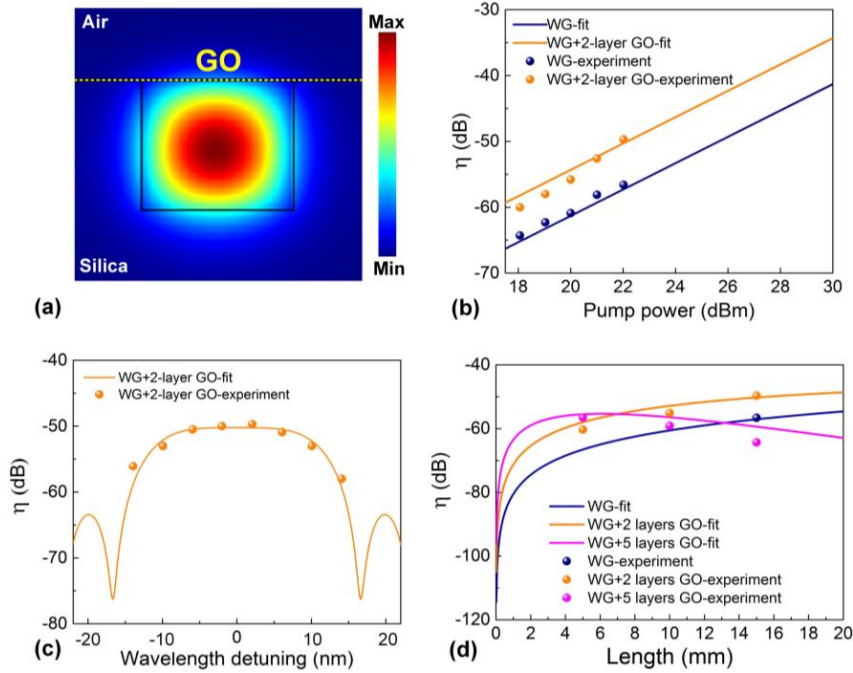

Fig. 4. (a) TE mode profile ($E_x$) of the hybrid doped silica waveguide with 2 layers of GO at a wavelength of 1550 nm. (b) − (d) $\eta$ as a function of pump power, wavelength detuning, and GO lengths, respectively. The dots represent the experimentally measured values and the lines show the fit curve calculated based on Eqs. (3) − (5). WG: waveguide.

Our values of $n_2$ are about a factor of ~5 lower than that reported in Ref. 34. This is perhaps not surprising since those measurements were performed at 800 nm where $n_2$ is expected[51–56] to be higher. Our value is also moderately higher than measurements reported on $n_2$ in the telecommunications band[35] and we suggest that this could be a result of our GO films being much thinner (≤ 10 nm) than those used in Ref. 35 (≥ 1 μm). Our experimentally measured FWM efficiency ($\eta$) in Figs. 4(b) – (d) agrees well with the curves obtained from Eqs. (3) – (5). For the dope silica waveguides studied here, Figure 4(d) shows a maximum theoretical improvement of ~20 dB, for 5 layers of GO and <1mm lengths to balance the trade-off between the FWM enhancement and propagation loss. By re-designing the waveguide cross-section to improve mode overlap with the GO film, we have calculated a theoretical enhancement in $\eta$ as large as 40 dB, again for short lengths (≤ 1 mm) and with 5 layers of GO films.

TABLE I. Comparison of $n_2$

|  | GO[34] | GO[35] | GO[a] |
|---|---|---|---|
| Wavelength (nm) | 800 | 1550 | 1550 |
| GO Film thickness | ~2 μm | ~1 μm | 2~10 nm |
| $n_2$ (m$^2$/W) | $7.5 \times 10^{-14}$ | $4.5 \times 10^{-15}$ | $1.5 \times 10^{-14}$ |
| $n_2$ normalized to silicon nitride[b] | $3 \times 10^5$ | $1.8 \times 10^4$ | $6 \times 10^4$ |

[a] Fit value according to the FWM experiment in this work.

[b] $n_2$ for silicon nitride = $2.5 \times 10^{-19}$ m$^2$/W (Refs. 3 and 31).

We use the values of $n_2$ for the GO films obtained from our FWM measurements as the basis for theoretical calculations of the FWM performance of silicon nanowires and slot waveguides integrated with GO films in order to demonstrate the universality of our approach. Recently, we achieved GO layers conformally coated on silicon nanostructures,[42] which is a good basis for the fabrication of these types of silicon-GO hybrid waveguides. Figure 5(a) shows the cross-sectional geometries and mode profiles of three typical waveguides in silicon-on-insulator (SOI) incorporated with GO. In our simulations, the TPA coefficient of silicon and the linear propagation loss of the silicon waveguides without GO were set to $5 \times 10^{-11}$ m$^2$/W[50] and 4.4 dB/cm,[57,58,59] respectively. The calculated $\gamma$'s for WWG-I, WWG-II, and SWG without GO and with 2 layers of GO are shown in Table II. Figure 5(b) − (d) depict the simulated FWM efficiency $\eta$ as a function of pump power, wavelength detuning, and waveguide length for the three silicon waveguides without GO and with 2 layers of GO. Compared to the silicon waveguides without GO, the maximum improvement in $\eta$ for the thin nanowire waveguide (WWG-II)[59] and slot waveguide[22] incorporated with GO is more than 30 dB. We note that the difference between using Eq. (10) and the more simplified Eq. (11) relating $\gamma$ and $n_2$ was only about 20% for the doped silica waveguides, but relatively larger for the silicon waveguides because of their much higher index contrast.

TABLE II. Calculated $\gamma$ for GO-silicon hybrid waveguides[a]

|  | WWG-I | WWG-II | SWG |
|---|---|---|---|
| $\gamma$ (W$^{-1}$m$^{-1}$) SOI waveguides | 255.0 | 71.4 | 96.0 |
| $\gamma$ (W$^{-1}$m$^{-1}$) SOI + GO | $1.14 \times 10^3$ | $3.23 \times 10^3$ | $1.65 \times 10^4$ |
| Ratio | 5.1 | 45.2 | 172 |

[a] $\gamma$ calculated using Eq. (10) and the $n_2$ for the GO films obtained from the FWM experiments.

The nonlinear figure of merit (FOM = $n_2/\beta_{TPA}$) [60] is a key factor in determining the efficiency of the Kerr nonlinear process, and under the conditions where Eq. (1) is valid, this also applies to FWM. Previously, [34] in Z-scan measurements of relatively thick GO films we showed that the FOM of GO was ~1 in unprocessed films measured at ~800 nm. Subsequently, $n_2$ was measured near 1550 nm and the FOM was found to be ~0.5, also in thick films. In this work, we did not observe any effects due to TPA and so were not able to estimate the nonlinear FOM, except to say that its effects were negligible.

We note that the concept of the nonlinear FOM was originally proposed[60] purely in the context of a limitation for $n_2$ devices having a positive $n_2$ and TPA. However, GO films have been demonstrated to exhibit very complex behavior (particularly in the context of photoreduction – see below), including both a negative $n_2$ as well as negative nonlinear absorption (e.g., due to saturable absorption), and it is not clear that the conventional FOM concept is relevant or useful in these cases. It is possible that for some of these cases, TPA may have a beneficial effect as has been demonstrated before[61, 62]. Finally, GO films are expected to similarly show significant enhancement in other nonlinear processes such as harmonic generation in photonic crystal waveguides[63-65], quantum effects relying on nonlinear processes[66], photonic crystals[67, 68] and even in long waveguides where dispersion is a key factor[69].

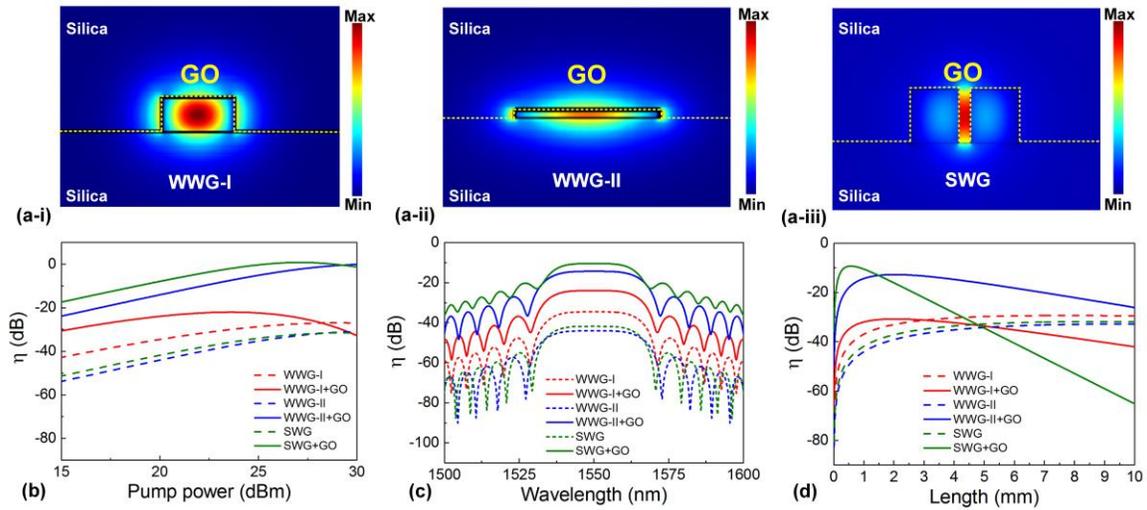

Fig. 5. (a) TE mode profile ($E_x$) of the silicon waveguides incorporated with 2 layers of GO at a wavelength of 1550 nm. (i) Wire waveguide-I (WWG-I) with a cross section of 500 nm × 220 nm. (ii) WWG-II with a cross section of 800 nm × 60 nm. (iii) Slot waveguide (SWG) with a slot width of 50 nm. The height and width of the silicon ribs are 220 nm and 200 nm, respectively. (b) − (d) Simulated $\eta$ as a function of pump power, wavelength detuning, and waveguide length for the silicon waveguides in (a), respectively. The pump wavelength in (b) − (d) is 1550 nm. The waveguide length in (b) and (c) is 1 mm. The pump power in (c) and (d) is 20 dBm. The signal wavelength in (b) and (d) is 1549 nm.

Finally, as we showed previously,[34] the material properties of GO can be changed by laser-induced photo-reduction processes, which can induce a permanent change in both $n_2$ and the FOM over a continuous wide range including even negative $n_2$ and negative nonlinear absorption (saturable absorption). This opens up the very powerful possibility of engineering structures to achieve quasi-phase matching, using the uniform negative $n_2$ in normal dispersion waveguides, and even using the photosensitivity to directly write cavities and waveguides, as has been done in chalcogenide photonic crystals[66-72]. Note however, that this process induces permanent material changes in the GO film and so is distinct from the super-quadratic power dependent behavior discussed above. Hence the potential of GO films for enhancing nonlinear processes in waveguides and nanowires extends well beyond the simple enhancement in overall nonlinear efficiency reported here.

## VI. CONCLUSION

We perform FWM measurements in doped silica waveguides integrated with thin GO films. We achieve a significant enhancement in the FWM conversion efficiency of ~9.5 dB in a 1.5-cm-long waveguide with 2 layers of GO. This results from the high Kerr nonlinearity, low linear loss, and the strong mode overlap of the GO films. The value of $n_2$ that we extract from our measurements agrees reasonably well with our previous Z-scan measurements of thick ($\geq 1$ μm) films. Further, we show theoretically that the enhancement in the FWM efficiency through the integration of thin GO films can be as high as 20 dB by optimizing the doped silica waveguides studied here, and is >30 dB in silicon thin nanowire and slot waveguides. With the potential for photo-patterning the nonlinearity of the GO films, these hybrid integrated devices offer a powerful new way to implement high performance nonlinear photonic devices, thus holding great promise for future ultra-high-speed all-optical information processing.

ACKNOWLEDGMENTS

This work was supported by the Australian Research Council Discovery Projects Program (No. DP150102972 and DP150104327). RM acknowledges support by Natural Sciences and Engineering Research Council of Canada (NSERC) through the Strategic, Discovery and Acceleration Grants Schemes, by the MESI PSR-SIIRI Initiative in Quebec, and by the Canada Research Chair Program. He also acknowledges additional support by the Government of the Russian Federation through the ITMO Fellowship and Professorship Program (grant 074-U 01) and by the 1000 Talents Sichuan Program in China.